\documentclass[12pt]{article}

\usepackage[english]{babel}
\usepackage{lmodern}
\usepackage{a4wide}
\usepackage{hyperref}
\usepackage{enumerate}

\usepackage{amssymb, amsmath, amsthm, dsfont}
\usepackage[matrix, arrow]{xy}
\usepackage[parsep]{collref}

\renewcommand{\d}{\ensuremath{\textnormal{d}}}

\newcommand{\id}{\mathds{1}}
\newcommand{\del}{\partial}
\def\d {{\rm d}}

\def\lll{{\cal L}}

\def\eee{{\cal E}}
\def\hhh{{\cal H}}

\def\teee{\tilde{{\cal E}}}
\def\tL{\tilde{{\cal L}}}
\def\te{\tilde{e}}

\def\tg{\tilde{g}}

\def\tp {\tilde{\phi}}

\def\tb {\tilde{\beta}}

\def\he{\hat{e}}
\def\hg{\hat{g}}
\def\hB {\hat{B}}
\def\hH {\hat{H}}
\def\hp {\hat{\phi}}
\def\heee{\hat{{\cal E}}}
\def\hL{\hat{{\cal L}}}

\def\R {\mathcal{R}}
\def\tR{\widetilde{\mathcal{R}}}
\def\hR{\widehat{\mathcal{R}}}

\begin{document}

\begin{titlepage}

\rightline{\small LMU-ASC 18/12}
\rightline{\small MPP-2012-64}

\vskip 2.8cm 

{\fontsize{18.2}{21}\selectfont 
  \flushleft{\noindent\textbf{Non-geometric $Q$-flux in ten dimensions\footnote{Contribution to the Proceedings of the Corfu Summer Institute 2011 "School and Workshops on Elementary Particle Physics and Gravity", September 4-18, 2011; Corfu, Greece}}} }

\vskip 0.2cm
\noindent\rule[1ex]{\textwidth}{1pt}
\vskip 1.3cm

\noindent\textbf{Peter Patalong$^{a,b}$}

\vskip 0.6cm
\begin{enumerate}[$^a$]
\item \textit{Arnold-Sommerfeld-Center for Theoretical Physics\\Department f\"ur Physik, Ludwig-Maximilians-Universit\"at M\"unchen\\Theresienstra\ss e 37, 80333 M\"unchen, Germany}
\vskip 0.2cm
\item \textit{Max-Planck-Institut f\"ur Physik\\F\"ohringer Ring 6, 80805 M\"unchen, Germany}
\end{enumerate}
%
\noindent {\small{\texttt{peter.patalong@physik.uni-muenchen.de}}}

\vskip 2.5cm

\begin{center}
{\bf Abstract}
\end{center}

\noindent We review the main ideas of \cite{allp11}: A change of field variables in the NSNS Lagrangian of ten-dimensional supergravity reveals the non-geometric $Q$-flux expressed in terms of an antisymmetric bivector field. After a simple dimensional reduction, the corresponding term gives the known contribution in the four-dimensional potential. Problematic non-trivial monodromies of non-geometric backgrounds in ten dimensions can be absorbed in a total derivative term such that a well-defined Lagrangian is obtained.

\vfill

\end{titlepage}

\section{Introduction}
String theory features symmetries that do not have any analogue in point particle quantum field theories. T-duality, as one of these, relates string backgrounds with different geometries and furthermore suggests the existence of so-called non-geometries where the notion of spacetime as a continuous manifold breaks down. Non-geometry appears in two different but intimately related aspects:
\begin{itemize}
\item {\bf Non-geometric backgrounds} can be obtained as T-duals of consistent string theory backgrounds. They exceed the framework of ordinary geometry, as the transition functions between coordinate patches have to include T-duality transformations to make fields like the metric or the dilaton single-valued. Nonetheless, non-geometric backgrounds can be argued to be consistent string backgrounds by considering the underlying conformal field theory \cite{h04}.
\item {\bf Non-geometric fluxes} arise when one tries to find higher-dimensional lifts of gauged supergravities in four dimensions and can be viewed as duals of the well-known geometric fluxes $H$ and $f$ under T-duality \cite{stw05}:
\begin{equation}\label{scheme}
H_{abc} \rightarrow {f^a}_{bc} \rightarrow {Q_c}^{ab} \rightarrow R^{abc} \ .
\end{equation}
In particular, the new fluxes $Q$ and $R$ are necessary to formulate T-duality invariant superpotentials and gauge algebras in four dimensions.
\end{itemize}
These two aspects are related by the double role of T-duality: in ten dimensions it leads to the existence of non-geometric backgrounds, whereas in four dimensions it connects the different types of fluxes. Due to this double role it is natural to expect that non-geometric backgrounds can be identified through corresponding non-geometric fluxes.

From a purely four-dimensional reasoning, non-geometric backgrounds provide new features to stabilise moduli \cite{stw06} and to obtain de Sitter spacetimes \cite{cgm09b}. This turns non-geometric flux compactifications into a very interesting and promising area of research.

An important step in finding a framework which can naturally incorporate non-geometric constructions was made with the development of so-called doubled formalisms. As T-duality mixes momentum and winding modes,
the doubling of the number of coordinates or the dimension of the tangent space was proposed to help accommodating the stringy extra degrees of freedom. And indeed, it was possible to construct T-duality invariant formalisms
of which the two main representatives are double field theory \cite{hhz10} and generalised geometry \cite{csw11}. It has been suggested that these can help to describe geometric and non-geometric backgrounds in a unified way.

Unfortunately, it has so far been obscure how to formulate a ten-dimensional theory that contains non-geometric fluxes and in particular, no supergravity action that reduces to known four-dimensional behaviour was available. We provide some progress in this direction.




\section{Making non-geometric fluxes appear}
This article reviews \cite{allp11}, where we extended the scope of the ten-dimensional NSNS action by introducing non-geometric fluxes and drew a link to their four-dimensional counterparts. Our conjecture is that the $Q$-flux can be revealed as a quadratic term, similar to the Kalb-Ramond field strength. To this end, we perform a field redefinition in the NSNS Lagrangian which is inspired by generalised complex geometry. It connects $Q$ to the bivector $\beta$ which has appeared in various contexts in the literature and sometimes has been interpreted as a sign of non-geometry.

Generalised complex geometry provides an object that embeds T-duality transformations in a particularly convenient way. It is the generalised metric $\hhh$, given in terms of the metric and the Kalb-Ramond field as
\begin{equation}\label{genmetric}
\hhh = \begin{pmatrix} \hg-\hB \hg^{-1}\hB && \hB \hg^{-1} \\ -\hg^{-1} \hB && \hg^{-1}\end{pmatrix}\ .
\end{equation}
A T-duality on these fields can be performed by conjugating the generalised metric with an $O(d,d)$ matrix \cite{gpr94},
\begin{equation}
\hhh' = O^T\hhh O\ .
\end{equation}
This transformation property has helped to find the above mentioned T-duality covariant formulations of the NSNS action. In our case, the most interesting property of $\hhh$ is that it can be rewritten in terms of generalised vielbeins $\eee$:
\begin{equation}
\hhh = \eee^T \id_{2d} \eee\ .
\end{equation}
These vielbeins are not uniquely defined and may be changed by $O(d)\times O(d)$ transformations without altering the generalised metric itself. It can be shown that there are two particular frames where the generalised vielbeins are triangular \cite{gmpw08},
\begin{equation}
\heee=\begin{pmatrix} \he && 0 \\ -\he^{-T}\hB && \he^{-T} \end{pmatrix} \  , \quad \teee=\begin{pmatrix} \te && \te \tb \\ 0 && \te^{-T} \end{pmatrix} \ .
\end{equation}
Here, $e$ represents an ordinary vielbein, $e^T e=g$, and $\beta$ is an antisymmetric bivector. We denote the two field bases with a hat and a tilde, respectively. Writing out the generalised metric relates them via
\begin{equation}
\begin{pmatrix} \hg-\hB \hg^{-1}\hB && \hB \hg^{-1} \\ -\hg^{-1} \hB && \hg^{-1}\end{pmatrix} = \heee^T \id_{2d}\ \heee = \teee^T \id_{2d}\ \teee 
= \begin{pmatrix} \tg && \tg\tb \\ -\tb \tg && \tg^{-1}-\tb \tg \tb \end{pmatrix}\ .
\end{equation}
In particular, the metric and the Kalb-Ramond field can be expressed as
\begin{align}
\hg_{mn} &= (G^{-1}_{\pm})_{mk} \tg^{kp} (G^{-1}_{\mp})_{pn}\\
\hB_{mn} &= (G^{-1}_{\pm})_{mk} \tb^{kp} (G^{-1}_{\mp})_{pn} \ ,
\end{align}
where we used the abbreviation
\begin{equation}
G_{\pm}^{mn}=\tg^{mn}\pm\tb^{mn}\ .
\end{equation}
In addition, we define a new dilaton $\tp$ such that the measure of the action integral remains invariant:
\begin{equation}
e^{-2\tp} \sqrt{|\tg|}=e^{-2\hp} \sqrt{|\hg|}\ .
\end{equation}
With these few expressions at hand, it is possible to rewrite the NSNS Lagrangian,
\begin{equation}\label{eq:lagrangianH}
\hL= e^{-2\hp} \sqrt{|\hg|} \left(\hR + 4|\d \hp|^2 - \frac{1}{2} |\hH|^2 \right)\ ,
\end{equation}
by simply plugging them in. This can be done in full generality \cite{ahllp12}, but we will focus on the special case where we employ a simplifying assumption,
\begin{equation}\label{assumption}
\tb^{mn}\del_n=0\ .
\end{equation}
It turns out, that this assumption basically corresponds to having no $R$-flux. Therefore, its simplifying power extends over pure technical aspects, as such an $R$-flux has been assigned to spoiling any locally geometric description by undermining the notion of points in spacetime \cite{gs06, l10, bp10, bdlpr11, cfl12}. Assuming \eqref{assumption} can thus be motivated by the avoidance of these problems.

It should be noted that performing the field redefinition requires an integration by parts in the scalar curvature term in order to remove terms with double derivatives on $\tg$ and $G^{-1}$. After a rather involved calculation, it is possible to get the overall result into the form
\begin{equation}\label{eq:lagrangianQ}
\boxed{\tL = e^{-2\tp} \sqrt{|\tg|} \left(\tR + 4|\d \tp|^2 - \frac{1}{2} |Q|^2 \right) \ ,}
\end{equation}
where the non-geometric flux $Q$ is defined as
\begin{equation}
{Q_m}^{np} = \partial_m \beta^{np} \ ,
\end{equation}
in agreement with other proposed definitions of $Q$ like for example in \cite{gmpw08}. This is a quite non-trivial result as interestingly all terms containing the unwieldy factors of $G^{-1}$ cancel out among themselves. Additionally, the remaining terms can be organised into the square of $Q$.

Before stating an equality between the two Lagrangians \eqref{eq:lagrangianH} and \eqref{eq:lagrangianQ}, one has to be careful about the total derivative we just mentioned. In non-geometric setups, it might turn out that total derivatives do not integrate to zero as the fields they contain may have non-trivial monodromies. Taking this into account we carefully state the refined result that
\begin{equation}
\boxed{\hL = \tL + (\text{total derivative}) \ .}
\end{equation}

These findings can be exemplified by the well-known $T^2$ fibration over $S^1$ with $H$-flux. It admits two T-dualities by symmetry, so that a situation with $Q$-flux can be obtained. The corresponding Lagrangian $\hL$ is ill-defined due to monodromies of the metric, but going to the redefined fields and the Lagrangian $\tL$ cures this problem and provides a well-defined action. Nevertheless this comes with a trade-off, as now the total derivative term turns out to be ill-defined and thus remains non-zero after integration. It was therefore argued in \cite{allp11}, as a kind of prescription, to neglect the total derivative and to describe the theory by only using the Lagrangian $\tL$. Furthermore, we argue that this prescription remains useful in more involved backgrounds as well.
%
\section{Compactification}
As a first check that the ten-dimensional $Q$-flux defined here can indeed be viewed as a proper lift of the four-dimensional non-geometric flux $Q$, we consider its scaling properties under fluctuations of the metric and the dilaton after a simple dimensional reduction. To this end, we define two moduli $\sigma$ and $\rho$ as four-dimensional fluctuations around their respective vacuum expectation value:
\begin{align}
g_{np}(x^m) \; &\rightarrow \; g_{np} \rho(x^\mu) \\
e^{-\phi(x^m)} \; &\rightarrow \; e^{-\phi} e^{-\varphi(x^\mu)} = g_s^{-1} e^{-\varphi(x^\mu)} \\
\sigma &= \rho^{3/2} e^{-\varphi} \ ,
\end{align}
where indices $m$ are ten-dimensional, and indices $\mu$ are four-dimensional. Performing the reduction leaves us with the four-dimensional action in Einstein frame,
\begin{equation}
S_E = M_4^2 \int \d^4 x \sqrt{|g_{\mu\nu}|} \left( \R_4 + \text{kin} + \sigma^{-2} \rho^{-1} \R_6 - \frac{1}{2} \sigma^{-2} \rho |Q|^2 \right) \ ,
\end{equation}
where $M_4$ denotes the four-dimensional Planck mass $M_4^2=L_0^6/(2\kappa^2g_s^2)$ with $L_0^6$ the internal volume, and kin denotes the kinetic terms for the moduli. The internal curvature scalar $\R_6$ and the $Q$-flux now denote vacuum expectation values. Comparing their scaling with respect to the defined moduli,
\begin{equation}
V_\omega \sim \sigma^{-2} \rho^{-1}\ ,\quad V_Q \sim \sigma^{-2} \rho\ ,
\end{equation}
we find full agreement with \cite{hktt08}, where the authors argued from a four-dimensional perspective.
%
\section{Relation to T-duality invariant formalisms}
Double field theory \cite{hhz10} provides an action that is manifestly T-duality invariant at the cost of doubling the coordinates. It makes use of the generalised metric \eqref{genmetric} and may thus be connected to the ideas presented here. In fact, our field redefinition leaves the generalised metric invariant and therefore corresponds to a particular $O(d,d)$ transformation in double field theory.

It is well-known that the double field theory action can be reduced to the standard NSNS action by simply taking the fields to be independent of the dual coordinates. We propose that the same reduction reproduces our $\tL$ when performed after applying the field redefinition in the form of a global $O(d,d)$ transformation. This can be summarised in the following diagram.
\begin{equation}
\xymatrix{ \lll_{DFT}(\hhh(\hg, \hB), \hat{\phi})\ \ar@{=}[r] \ar@{->}[d] & \ \lll_{DFT}(\hhh(\tg, \tb), \tilde{\phi}) \ar@{->}[d] \\
\hL+ \del(\dots)\ \ar@{=}[r] &\ \tL+ \del(\dots) }
\end{equation}
Recently, we were able to prove this proposal and found an interesting interpretation of the $Q$-flux as part of a connection that makes the winding derivatives covariant \cite{ahllp12}. Furthermore, it was possible to drop the simplifying assumption and reveal the $R$-flux in its fully covariant form.
%
\section{Outlook}
\noindent There are several directions into which this work can be extended.
\begin{itemize}
\item Here, we only investigated the NSNS sector and it would be interesting to extend our ideas in a way that RR fluxes can be included as well. This may especially help to restore S-duality and points into the direction of exceptional generalised geometry.
\item As it has turned out that double field theory provides a framework where the full set of non-geometric fluxes can be described, it might be interesting to work out examples that go beyond the doubled torus with H-flux. Especially, it has to be clarified how the strong constraint of double field theory relates to the global obstructions of non-geometric backgrounds.
\item So far, the connection to four dimensions has been kept on a very basic level. It is of course desirable to draw the link between ten-dimensional non-geometric fluxes and the corresponding gauge algebra of the gauged supergravity in four dimensions. Some considerations along these lines can for example be found in \cite{abmn11, g11}.
\end{itemize}
%
\section*{Acknowledgements}
It is my pleasure to thank the organisers of the Corfu Summer Institute 2011 for giving me the opportunity to speak at the ``Workshop on Fields and Strings''. I also want to thank M. Larfors and D. L\"ust for valuable comments during the preparation of this article.

\bibliographystyle{JHEP}
\providecommand{\href}[2]{#2}\begingroup\raggedright\endgroup

\end{document}